# An effective medium theory for predicting the existence of surface states

Meng Xiao, Xueqin Huang, Anan Fang, C. T. Chan[*]

Department of Physics and Institute for Advanced Study, Hong Kong University of Science and Technology, Clear Water Bay, Kowloon, Hong Kong, China

[*]Corresponding author: phchan@ust.hk

Abstract: We build an effective medium theory for two-dimensional photonic crystals comprising a rectangular lattice of dielectric cylinders with the incident electric field polarized along the axis of the cylinders. In particular, we discuss the feasibility of constructing an effective medium theory for the case where the Bloch wave vector is far away from the center of Brillouin zone, where the optical response of the photonic crystal is necessarily anisotropic and hence the effective medium description becomes inevitability angle dependent. We employ the scattering theory and treat the two-dimensional system as a stack of one-dimensional arrays. We consider only the zero-order interlayer diffraction and all the higher order diffraction terms of interlayer scattering are ignored. This approximation works well when the higher order diffraction terms are all evanescent waves and the interlayer distance is far enough for them to decay out. Scattering theory enables the calculation of transmission and reflection coefficients of a finite sized slab, and we extract the effective parameters such as the effective impedance ($Z_e$) and the effective refractive index ($n_e$) using a parameter retrieval method. We note that $n_e$ is uniquely defined only in a very limited region of the reciprocal space. ($n_e k_0 a \ll 1$, where $k_0$ is the wave vector inside the vacuum and $a$ is thickness of the slab for retrieval), but $Z_e$ is uniquely defined and has a well-defined meaning inside a much larger domain in the reciprocal space. For a lossless system, the effective impedance $Z_e$ is purely real for the pass band and purely imaginary in the band gaps. Using the sign of the imaginary part of $Z_e$, we can classify the band gaps into two groups and this classification explains why there is usually no surface state on the boundary of typical fully gapped photonic crystals comprising of a lattice of dielectric cylinders. This effective medium approach also allows us to predict the dispersion of surface states even when the surface wave vectors are well beyond the zone center.

# I Introduction

Photonic crystals (PCs) and metamaterials provide useful platforms for controlling the propagation of electromagnetic waves[1-3]. Due to the artificial periodicity of PCs and most of metamaterials, the electromagnetic responses of them are best described using the language of band structure or similar concepts. However, it is highly desirable to have an effective medium description in the long wave length limit and various methodologies have been developed for extracting effective parameters such as the electric permittivity and magnetic permeability from these otherwise complex and artificially structured materials. [4-15] One obvious example of the usefulness of accurate effective medium parameters is the determination of the existence surface or interface states and their dispersions without the need of going through tedious calculations.

In fact, the dispersion of interface states at the boundary separating two pieces of homogenous materials can be predicted analytically once the effective parameters are specified. However, artificially structural materials such as photonic crystals and metamaterials are inherently inhomogeneous and standard effective medium parameters would have difficulty accounting for interface states that extends to high values of *k* away from the center of Brillouin zone (BZ). This is because at high values of *k*, the wave can probe the details the microstructure and standard effective medium approach may fail. As conventional effective medium theories (EMT) such as those based on the coherent potential approximation (CPA)[7,8] require the condition $k_{eff} r_0 \ll 1$ ($k_{eff}$ is the wave vector in the effective media and $r_0$ represents the linear size of microstructure), they are not expected to provide a good approximation of a composite material if $k_{eff}$ is big in which case the effective wavelength inside the material is small enough to know the details of the microstructure. If we still insist on using effective parameters, such effective parameters should depend on the direction of incident. In other words, they should be spatially dispersive. If one employs the layer Korringa-Kohn-Rostoker (KKR) method[13-15], one can regard the PC or metamaterial as a stack of layers parallel to a given crystallographic plane of them. In that case, it is clear that besides frequency, the extracted effective parameters should also depend on the parallel component of the wave vector ($\vec{k}_p$) along the given crystallographic plane. In other words, the EMT becomes naturally angular ($\vec{k}_p$) dependent.

Let us now focus on a PC comprising a rectangular lattice of dielectric cylinders (the relative permeability equals one) embedded in the air. Following the paradigm of the KKR method[13-15], we treat the PC as a stack of layers parallel to a given crystallographic plane. To describe the scattering properties of this PC, the scattering field is first expanded as a summation of cylindrical functions and then transformed to linear combination of plane waves with different $\vec{k}_p + \vec{g}$, i.e.,

$$\vec{K}_{\vec{g}}^{\pm} = \left( \vec{k}_p + \vec{g}, \pm\sqrt{k^2 - \left\|\vec{k}_p + \vec{g}\right\|^2} \right), \quad (1)$$

where $\vec{g}$ is the reciprocal vector parallel to the chosen crystallographic plane, $\vec{k}_p$ is the Bloch wave vector parallel to the plane, and $\sqrt{k^2 - \left\|\vec{k}_p + \vec{g}\right\|^2}$ is the component of the plane wave normal to plane. The intra-layer coupling can be calculated in full rigor using standard lattice sum techniques. However, in order to describe the optical properties this system within the context of effective medium approach, we need to discard the interlayer interactions with reciprocal vector $\vec{g} \neq 0$. This approximation is applicable when the $\sqrt{k^2 - \left\|\vec{k}_p + \vec{g}\right\|^2}$ terms are all imaginary with reciprocal vector $\vec{g} \neq 0$ are all evanescent waves and at the same time, the interlayer distance is long enough for them to decay. This discarding $\vec{g} \neq 0$ approximation (no higher order diffraction terms) is effectively a homogenization process so that each layer of cylinders behave collectively as a homogenous slab as far as wave scattering is concerned. To extract the effective parameters which describe the homogenized slab, we can first calculate the reflection coefficient ($r$) and the transmission coefficient ($t$) with the KKR method. The effective parameters can then be obtained with the retrieval methods [9-11] based on requirement the effective parameters gives the same scattering property (same $r$ and $t$) as the real system. We will call this apporach "layer EMT" and we will show in the following that this layer EMT can enable us to extend to the effective region of the EMT away from the $\Gamma$ point (i.e., $k_{eff} r_0$ being small is not a necessary requirement anymore) and thereby allowing us to predict the dispersion of surface waves with large values of $\vec{k}_p$.

The paper is organized as follows: In Section II, we discuss in detail how to construct the layer EMT using a layer-by-layer scattering theory and a retrieval method to extract the effective parameters of the system. In Section III, we compare this EMT with some conventional EMTs, including the CPA based method[7], the boundary EMT[16, 17] and retrieval method[9-11]. We also compare the projected band structure as well as the transmission spectrum of a slab of finite thickness predicted with the layer

EMT and those from full wave simulation. In Section IV, we discuss some of the possible applications of this EMT. In particular, we will use this approach to explain why there is usually no interface state on the boundary of typical fully gapped PCs comprising of dielectric cylinders. We also show that our EMT can predict the dispersions of surface states between a PC and a homogenous material. In Section V, we discuss some possible extensions and potential applications of the layer EMT and then give our conclusions in Section VI.

## II Formulation

Our system is shown schematically in Fig. 1(a), which consists of an array of identical dielectric cylinders (blue disks) forming a rectangular lattice embedded in a vacuum and the lattice constant is $a$ ($b$) along the $x$- ($y$-) direction. The radius, the relative permittivity, permeability and refractive index of the cylinder are given by $r_c$, $\varepsilon_c$, $\mu_c$ and $n_c$, respectively, and $n_c = \sqrt{\varepsilon_c \mu_c}$. In our numerical simulations we use non-magnetic cylinder with $\mu_c = 1$, and our theory also applies when $\mu_c$ does not equal one. A plane wave with the electric field parallel to the axis of the cylinder (the $z$ direction) impinges on this PC. The wave vector of this plane wave is given by $\vec{k}_0$ and the component parallel to the $y$ direction is $k_p$. The incident plane wave is hence specified as

$$\vec{E}^{inc} = E_0 \exp\left[i\left(k_p y + k_x x\right)\right]\hat{z}, \text{ (2a)}$$

$$H_x^{inc} = \frac{k_p}{\omega \mu_0} E_0 \exp\left[i\left(k_p y + k_x x\right)\right], \text{ (2b)}$$

$$H_y^{inc} = -\frac{k_x}{\omega \mu_0} E_0 \exp\left[i\left(k_p y + k_x x\right)\right], \text{ (2c)}$$

where $\omega$ is the angular frequency and $k_x = \sqrt{k_0^2 - k_p^2}$ is the wave vector along the $x$ direction. This two-dimensional (2D) PC can be treated as a stack of one-dimensional (1D) layers with virtual boundaries marked with dashed yellow lines in Fig. 1(a). To study the properties of this 2D PC, we start with a single layer of such cylinders, and calculate the transmission coefficient ($t$) and reflection coefficient ($r$) at the virtual boundaries (indicated with the dashed yellow lines in Fig. 1(b)) of this layer. The layer is chosen in a central symmetric manner with reference to the center of the cylinders. The lowest three modes of a dielectric cylinder in a vacuum consists of one electric monopole ($P_z$) and two magnetic dipoles ($M_x$ and $M_y$, which represent magnetic dipoles along the $x$ and $y$ direction, respectively.), and these three modes determine the scattering properties of the cylinder when the frequency of the incident wave is not too high. The excitation strength of the monopole and dipoles are given by[18, 19]

$$P_z = \alpha_E E_z^{loc}, \quad (3a)$$

$$\vec{M} = \alpha_M \vec{H}^{loc}, \quad (3b)$$

where $E_z^{loc}$ and $\vec{H}^{loc}$ are the $z$ component of the local electric field and the local magnetic field at the position of the cylinder, and $\alpha_E$ and $\alpha_M$ are the polarizability of monopole and dipole of a cylinder, respectively, which are given by

$$\alpha_E = \frac{4i\varepsilon_0}{k_0^2} \beta_0(r_c k_0), \quad (4a)$$

$$\alpha_M = \frac{8i}{k_0^2} \beta_1(r_c k_0), \quad (4b)$$

and

$$\beta_n(r_c k_0) = \frac{\mu_c J_n(n_c r_c k_0) J_n'(r_c k_0) - n_c J_n(r_c k_0) J_n'(n_c r_c k_0)}{\mu_c J_n(n_c r_c k_0) H_n'^{(1)}(r_c k_0) - n_c H_n^{(1)}(r_c k_0) J_n'(n_c r_c k_0)}, \quad (5)$$

where $J_n$, $H_n^{(1)}$, $J_n'$ and $H_n'^{(1)}$ are the $n^{th}$ order Bessel function, Hankel function of the first kind and their derivatives, respectively. The local fields consist of the incident plane wave field and the scattering fields from other cylinders, the monopole and magnetic dipoles of the $l^{th}$ cylinder are hence given by

$$P_l = \alpha_E \left\{ E_l^{inc} + \sum_{m \neq l} \left[ \frac{ik_0^2}{4\varepsilon_0} P_m H_0^{(1)}\left(k_0 |\vec{r} - \vec{r}_m|\right) + i\omega\mu_0 \hat{z} \cdot \nabla \times \left( \frac{i}{4} \vec{M}_m H_0^{(1)}\left(k_0 |\vec{r} - \vec{r}_m|\right) \right) \right] \Big|_{\vec{r}=\vec{r}_l} \right\}, \quad (6a)$$

$$\vec{M}_l = \alpha_M \left\{ \vec{H}_l^{inc} + \sum_{m \neq l} \left[ \nabla \times \left( \frac{\omega}{4} P_m H_0^{(1)}\left(k_0 |\vec{r} - \vec{r}_m|\right) \hat{z} \right) + \nabla \times \nabla \times \left( \frac{i}{4} \vec{M}_m H_0^{(1)}\left(k_0 |\vec{r} - \vec{r}_m|\right) \right) \right] \Big|_{\vec{r}=\vec{r}_l} \right\}, \quad (6b)$$

where $E_l^{inc}$ and $\vec{H}_l^{inc}$ are the electric field along the $z$-direction and magnetic field of the incident wave at the position of the $l^{th}$ cylinder, and $\vec{r}_l = (a/2, lb)$. As the system is periodic along the $y$ direction, the Bloch condition along the $y$ direction can be applied, and

$$P_m = \exp(imbk_p) P, \quad (7a)$$

$$\vec{M}_m = \exp(imbk_p) \vec{M}, \quad (7b)$$

where $P$ and $\vec{M}$ are the monopole and magnetic dipole moments of the $0^{th}$ cylinder. Take Eq. (7) into Eq. (6), we have

$$\begin{pmatrix} \dfrac{\varepsilon_0}{k_0^2 \alpha_E} - F_1 & -F_3 & 0 \\ F_3 & \dfrac{1}{\alpha_M} + F_6 & 0 \\ 0 & 0 & \dfrac{1}{\alpha_M} + F_4 \end{pmatrix} \begin{pmatrix} i\omega P \\ M_x \\ M_y \end{pmatrix} = \begin{pmatrix} \dfrac{i}{\omega\mu_0} E^{inc} \\ H_x^{inc} \\ H_y^{inc} \end{pmatrix}, \quad (8)$$

where $F_1$, $F_3$, $F_4$, $F_6$ are lattice sums defined as

$$F_1 = \sum_{m \neq 0} \frac{i}{4} H_0^{(1)}(mbk_0) e^{imk_p b}, \quad (9a)$$

$$F_3 = \frac{\partial}{\partial y} \sum_{m \neq 0} \frac{i}{4} H_0^{(1)}(mbk_0) e^{imk_p b} \Big|_{\vec{r}=0}, \quad (9b)$$

$$F_4 = \frac{\partial^2}{\partial x^2} \sum_{m \neq 0} \frac{i}{4} H_0^{(1)}(mbk_0) e^{imk_p b} \Big|_{\vec{r}=0}, \quad (9c)$$

$$F_6 = \frac{\partial^2}{\partial y^2} \sum_{m \neq 0} \frac{i}{4} H_0^{(1)}(mbk_0) e^{imk_p b} \Big|_{\vec{r}=0}. \quad (9d)$$

These lattice sums converge slowly in the real space, so we need to transform to the reciprocal space, and the results are as follows:[20]

$$F_1 = \frac{1}{2\pi}\left(\ln\left(\frac{k_0 b}{4\pi}\right) + \gamma_E\right) + i\left(\frac{1}{2bk_x} - \frac{1}{4}\right) + \frac{1}{2b}\sum_{m=1}^{\infty}\left(\frac{i}{q_m} + \frac{i}{q_{-m}} - \frac{2}{g_m}\right), \quad (10a)$$

$$F_3 = -\frac{k_p}{2bk_x} - i\frac{k_p}{2\pi}\sum_{m=1}^{\infty}\left[\frac{(k_p - g_m)}{2bq_m} + \frac{(k_p + g_m)}{2bq_{-m}}\right], \quad (10b)$$

$$F_4 = -\frac{k_0^2}{4\pi}\left(\ln\left(\frac{k_0 b}{4\pi}\right) + \gamma_E - \frac{1}{2}\right) - \frac{k_p^2}{4\pi} - \frac{\pi}{6b^2} + i\left(\frac{k_0^2}{8} - \frac{k_x}{2b}\right) - \frac{1}{2b}\sum_{m=1}^{\infty}\left(iq_m + iq_{-m} + 2g_m - \frac{k_0^2}{g_m}\right), \quad (10c)$$

$$F_6 = -\frac{k_0^2}{4\pi}\left(\ln\left(\frac{k_0 b}{4\pi}\right) + \gamma_E + \frac{1}{2}\right) + \frac{k_p^2}{4\pi} + \frac{\pi}{6b^2} + i\left(\frac{k_0^2}{8} - \frac{k_p^2}{2bk_x}\right) - \frac{1}{2b}\sum_{m=1}^{\infty}\left(\frac{i(k_p - g_m)^2}{q_m} + \frac{i(k_p + g_m)^2}{q_{-m}} - 2g_m - \frac{k_0^2}{g_m}\right),$$

(10d)

where $\gamma_E$ is the Euler constant, $g_m = 2\pi m/b$, $k_0^2 = q_m^2 + (k_p - g_m)^2$ ($m \in \mathbb{Z}$). According to Eq. (2), the incident fields at the center of the $0^{\text{th}}$ cylinder, i.e., $x = a/2$, $y = 0$, are $E^{inc} = E_0 \exp(ik_x a/2)$,

$H_x^{inc} = \frac{k_p}{\omega\mu_0} E_0 \exp(ik_x a/2)$ and $H_y^{inc} = -\frac{k_x}{\omega\mu_0} E_0 \exp(ik_x a/2)$. If we define

$\tilde{P} = \omega^2 \mu_0 P \exp(-ik_x a/2)/E_0$, $\tilde{M}_x = \omega\mu_0 M_x \exp(-ik_x a/2)/E_0$ and

$\tilde{M}_y = \omega\mu_0 M_y \exp(-ik_x a/2)/E_0$, then Eq. (8) can be written as

$$\begin{pmatrix} i\tilde{P} \\ \tilde{M}_x \\ \tilde{M}_y \end{pmatrix} = \begin{pmatrix} \frac{1}{\alpha_M} + F_6 & F_3 & 0 \\ -F_3 & \frac{\varepsilon_0}{k_0^2 \alpha_E} - F_1 & 0 \\ 0 & 0 & \frac{\alpha_M}{1 + F_4 \alpha_M} \end{pmatrix} \begin{pmatrix} i/\zeta \\ k_p/\zeta \\ -k_x \end{pmatrix}, \quad (11)$$

where $\zeta = \left(\frac{1}{\alpha_M} + F_6\right)\left(\frac{\varepsilon_0}{k_0^2 \alpha_E} - F_1\right) + F_3^2$. Up to this point, the only approximation is that we just keep the monopole and dipole excitations of the dielectric cylinder. The scattering field of this layer of cylinders is expanded as a summation of plane waves with the parallel component of the wave vector given by $k_p + g_m$. When all the diffraction terms with $g_m \neq 0$ are evanescent waves and the interlayer distance is large enough for these diffraction terms to decay out, the dominant component of the scattering field is the lowest order ($g_m = 0$) plane wave term. We shall ignore all the higher order diffraction terms when dealing with the interlayer scattering in order to build an EMT. We note here that we keep the complete summation in Eq. (10a) when we are calculating the intralayer interaction. This approximation works when $|k_p \pm 2\pi/b| > k_0$ in which case the $x$ component of the wave vector of the $m^{th}$ ($m \in \mathbb{Z}$) diffraction term with $g_m \neq 0$ is given by $\sqrt{k_0^2 - (k_p - 2m\pi/b)^2}$ and is purely imaginary. Close to the zone boundary $|k_p \pm 2\pi/b| = k_0$, the decay length (inversely proportional to the imaginary part of the wave vector along the $x$ direction) of the $\pm 1$ order diffraction term becomes quite large, and hence this approximation becomes poor. This is consistent with the fact of EMT approaches would typically fail when the wavevector approach the zone boundary. Under this approximation, the scattering field can be written as

$$\vec{E}_s = \hat{z}\frac{ik_0^2}{\varepsilon_0}\left(\frac{P}{2bk_x} + \frac{M_x k_p}{2b\omega k_x} - \frac{M_y \, \text{sgn}(x - a/2)}{2b\omega}\right)\exp(ik_x|x - a/2| + ik_p y), \quad (12)$$

and the reflection field and transmission field are

$$\vec{E}_r = \frac{iE_0}{2bk_x}\left(\tilde{P} + \tilde{M}_x k_p + \tilde{M}_y k_x\right)e^{i(k_x a + k_p y)}\hat{z} \quad (13)$$

$$\vec{E}_t = \left[1 + \frac{i}{2bk_x}\left(\tilde{P} + \tilde{M}_x k_p - \tilde{M}_y k_x\right)\right]E_0 e^{i(k_x a + k_p y)}\hat{z}. \quad (14)$$

Then we obtain the reflection and transmission coefficient at the virtual boundary in Fig. 1(b) as

$$r = \frac{i}{2bk_x}\left(\tilde{P} + \tilde{M}_x k_p + \tilde{M}_y k_x\right)e^{ik_x a}, \quad (15)$$

$$t = \left[1 + \frac{i}{2bk_x}\left(\tilde{P} + \tilde{M}_x k_p - \tilde{M}_y k_x\right)\right]e^{ik_x a}. \quad (16)$$

Up to now, we have solved the scattering problem of a single column of identical cylinders analytically. Following the approach of the retrieval methods[9-11], we look for a layer of homogenous material with thickness $a$ that has the same $r$ and $t$, and then solve for the effective parameters of the homogenous material. Here we intend to find the effective refractive index ($n_e$) and relative impedance ($Z_e$) instead of the effective relative permittivity and permeability usually given in retrieval methods. The relative effective impedance is defined as the ratio between the electric field along the z direction and the magnetic field along the y-direction, i.e., $Z_e = -E_z/(H_y Z_0)$, where $Z_0$ is the impedance of vacuum and the minus sign is used to ensure that $Z_e$ equals bulk impedance when $k_p = 0$. The value of effective impedance tells the reflection and transmission of a plane wave at a planar boundary between two different materials. $Z_e$ depends on how the PC is truncated. Different surfaces (different positions or directions) of a PC have different values of impedance, and hence we call it surface impedance. To be specific, the surface of a PC is specified to be along the y direction and has a distance of $a/2$ from the outmost column of cylinders. The effective refractive index tells the phase delay inside the PC. As we use the transmission and reflection coefficient at the boundary for retrieval, there is a well-known uncertainly of $2m\pi$ ($m \in \mathbb{Z}$) in the determination of the phase delay. After some mathematics, we obtain $n_e$ and $Z_e$ as functions of $\omega$ and $k_p$:

$$Z_e(\omega, k_p) = \pm \frac{\sqrt{(r+1)^2 - t^2}}{\sqrt{\varepsilon_0 - k_p^2/(k_0^2 \mu_0)} \sqrt{(r-1)^2 - t^2}}, \quad (17)$$

$$n_e(\omega, k_p) = \pm \frac{1}{k_0 a} \arccos\left(\frac{1 - r^2 + t^2}{2t}\right) + \frac{2\pi}{k_0 a} m \, (m \in \mathbb{Z}) \quad (18)$$

$$\tilde{n}_e(\omega, k_p) = \pm \sqrt{n_e^2 + k_p^2/k_0^2}. \quad (19)$$

Here $n_e$ records the phase delay along the $x$ direction and $\tilde{n}_e$ can be interpreted as the bulk refractive index in the low frequency limit. It can be proved that (see Appendix A) if there is no absorption or gain in the system, $Z_e$ is purely real or purely imaginary while $\cos(n_e k_0 a)$ is purely real in Eqs. (17) and (18). Let us consider a plane wave impinges on a semi-infinite PC and the reflection coefficient is given by *r*. From energy conservation consideration, $|r| \leq 1$, and this in turn requires $\text{Re}(Z_e) \geq 0$ since $r = (z-1)/(1+z)$. On the other hand, to ensure the amplitude of the transmitted wave is finite at infinity, $\text{Im}(n_e)$ must be positive. If $Z_e$ is purely imaginary, then $|r| = 1$ and inside the gap region. Otherwise, if $Z_e$ is purely real, it must be inside a passband and *vice versa*. There are, however, some discrete points inside the passband at which $|r|=1$, and $Z_e$ equals 0 or $\pm\infty$. If $|\cos(n_e k_0 a)| \leq 1$, $n_e$ is purely real and the corresponding frequency is inside the passband; if $|\cos(n_e k_0 a)| > 1$, the imaginary part of $n_e$ is not zero and the corresponding frequency is inside the bandgap and *vice versa*.

It can also be proved that (see Appendix B) these effective parameters retrieved from only one layer of cylinder are the same as those from several layers of cylinder once the unit cell chosen for retrieval has mirror symmetry relative to the center of the unit cell and the higher order interlayer diffraction terms can be ignored. In other words, the parameters are independent of the number of layer for retrieval, we can hence use the parameters retrieved from a single layer of cylinders to describe a bulk PC.

## III Comparison with other EMT methods

In the last section, we analytically derived an effective medium description for 2D PCs for an rectangular lattice. In this section, we will compare our layer EMT with conventional EMT theories including CPA-based method[7], boundary EMT[16, 17], parameter retrieval method[9-11, 21-23]. We will also compare with full wave simulations.

## A. The low frequency limit

Let us start with the low frequency limit, the linear dispersion region where both wave vector $k$ and angular frequency $\omega$ go to zero. Traditional EMTs[4-6] work in the limit of $k \to 0$ and $\omega \to 0$. More modern approaches such as CPA based method[7] and boundary EMT[16, 17] can extend the range of validity to higher frequencies and can take care of resonances, but they should of course give the same effective parameters when $k \to 0$ and $\omega \to 0$. In this part, we will compare our EMT (named layer EMT) with the CPA based method[7] and the boundary EMT[22, 23]. In Figs. 2(a) and 2(b), the effective parameters ($n_e$ and $Z_e$) obtained using one implementation of the CPA based method[7], an improved version of the Maxwell-Garnett method (open black circles), the boundary EMT[16, 17] (solid blue triangles) and the layer EMT (solid red line) are compared. In this calculation we consider a PC composing of a square lattice ($a=b$) of cylinders with $r_c = 0.2a$ embedded in a vacuum. We fix the angular frequency at $\omega = 0.02\pi c/a$, and vary the relative permittivity of the cylinder $\varepsilon_c$. Parameters are chosen such that, condition $n_e k_0 r_c \ll 1$ is satisfied as shown in the inset in Fig.2 (a), where we plot $n_e k_0 r_c$ as a function of $\varepsilon_c$ over the parameter range we considered. As $n_e k_0 r_c \ll 1$, the CPA method and the boundary EMT should work well which is also confirmed as these two methods show exactly the same results in Figs. 2(a) and 2(b). As said before, the layer EMT depends on the angle of incidence. For a meaningful comparison with other EMT methods, we consider the case $k_p = 0$ (normal incidence). The effective parameters obtained with our layer EMT in Figs. 2(a) and 2(b) show exactly the same results as the above mentioned two methods. So we can conclude that our layer EMT satisfies the condition of giving the same results as other formulations in the low frequency limit.

## B. Beyond the low frequency limit with $k_p = 0$

Next we go beyond the low frequency limit while still keep $k_p = 0$. We discuss the properties of the EMTs over a large frequency range where the condition $n_e k_0 r_c \ll 1$ is not guaranteed for some frequency ranges. In this case, The parameters of the square lattice PC are given by $\varepsilon_c = 10$, $r_c = 0.18a$. We study the frequency dependence of the effective parameters and then compare them with other methods.

First, let us start with the effective refractive index ($n_e$). To avoid choosing branch in Eq. (18), here we plot $\cos(n_e k_0 a)$ instead of $n_e$. The purple line, red line, green triangles and blue line in Fig. 3(b) represent the results calculated using the CPA based method[7], our layer EMT, the layer KKR method[13-15] and the boundary EMT[16, 17], respectively. Here we use Eq. (7) of Ref[7] when applying the CPA based method. Layer KKR method is only used inside the bang gap region as a complement to the boundary EMT as the boundary EMT cannot be applied inside the band gap (no eigenstate exists inside the band gap). In Fig. 3(b), the boundary of the region where $|\cos(n_e k_0 a)| \leq 1$ are marked by two black dashed lines. When $|\cos(n_e k_0 a)| > 1$, the imaginary part of $n_e$ is not zero, and this corresponds to frequency inside the band gap. We plot the band dispersion (solid black lines) along the $\Gamma X$ direction in Fig. 3(a), which can serve as criteria on whether one EMT behaviors well or not in predicting the band edge frequencies. From this point of view, our layer EMT gives good predictions on the band edge frequencies while the CPA method fails as the *k* vector moves away from the zone center. This is because the CPA formulae are derived under the assumption $|n_e k_0 a| \ll 1$ and this condition is not satisfied near the first band gap. The CPA method[7] and the boundary EMT[22, 23] show resonance behavior for the effective relative permittivity (permeability) at the lower (upper) edge of the first band gap. Near the resonance frequency, when the relative permittivity and permeability have the same sign, $n_e$ is a large real number and this cause $\cos(n_e k_0 a)$ to oscillate quickly near the band edge. This oscillation behavior comes from the artificially chosen boundary in the CPA method and the boundary EMT which does not exist in the full wave simulation. This oscillation behavior is also absent with our layer EMT method. For each frequency inside the highest pass band (around $\omega a/(2\pi c) = 0.7$) in Fig. 3(a) (also in Fig. 4(a)), there are two eigenstates for positive k corresponding to the quasi-longitudinal and transverse dipole modes[24]. Because the longitudinal dipole mode couples poorly with the incident wave, we keep only the transverse dipole mode to obtain the effective parameters when using the Boundary EMT. In Fig. 3(b), we choose not to show the results obtained from the CPA method and the Boundary EMT inside the band gaps for the following reasons. First, the Boundary EMT works only inside the pass band, because one needs to obtain the eigenfield distribution[22, 23] when applying the Boundary EMT and there is no state inside the band gap. Secondly, $n_e$ obtained from the CPA method is a large imaginary number inside the first band gap, and this means $|\cos(n_e k_0 a)|$ is huge inside the first band gap. The imaginary part of $n_e$ tells the decay length inside the band gaps, which can also be obtained using the layer KKR method. When applying layer KKR method, we obtain the eigenvalues of the transfer matrix[25] of one layer for each frequency and we only keep the branch which decays most slowly, i.e., the eigen wave vector with smallest complex part. The results are shown with green triangles in Fig. 3(b). Our layer EMT is consistent with the complex-K method inside the first band gap while deviates a little inside the second band gap. This deviation is due to the fact that we ignore all the higher order diffraction terms in our study and this approximation becomes progressively less accurate when the frequency is high enough.

Let us now discuss the effective impedance ($Z_e$). In Figs. 4(b) and 4(c), we show the real and the imaginary part of $Z_e$, respectively. As comparison, we also show the effective parameters obtained from the CPA method, the boundary EMT method and the retrieval method. As proved in the last section, $Z_e$ is purely imaginary inside the band gap and purely real inside the pass band, we plot the band dispersion (solid black line) along the $\Gamma X$ direction in Fig. 4(a) as a reference. The green background represents the pass band frequency range. The frequencies where the surface impedance change from purely real to purely imaginary or *vice versa* are consistent with the band edge frequencies. For the pass band regions, $Z_e$ obtained with the boundary EMT and the layer EMT are almost exactly the same while that obtained with the CPA method shows deviation close to the band edge. In particular, the CPA method finds that $Z_e$ is still a real number for some frequencies inside the first band gap (Fig. 4b). This is also because the CPA method works well only when $|n_e k_0 a| \ll 1$ and this condition is not always satisfied (see Fig. 3(b), $\cos(n_e k_0 a) \approx 1$ only in a limited frequency ranges). Since the Boundary EMT does not work inside the band gaps, we use a retrieval method to show the validity of $Z_e$ inside the band gaps. Since $Z_e$ tells the reflection phase of a semi-infinite system inside the band gap (reflection amplitude is one), we can calculate the reflection phase to verify our layer EMT. We calculate the reflection phase of a thick enough system (the majority of the incident wave is reflected back) inside the band gaps with full wave simulation using COMSOL, and then retrieve the surface impedance and the results are shown with green triangle. Our EMT agrees well with the results retrieved from full wave simulation.

## C. EMT for $k_p \neq 0$

Now we will show that this layer EMT can still be applied when $k_p \neq 0$. Here we just compare the projected band structures along the $\overline{\Gamma}\overline{X}$ direction calculated with full wave simulation and our EMT, and the corresponding results are shown in Figs. 5(a) and 5(b), respectively. The parameters of the square lattice PC are given by $\varepsilon_c = 10$, $r_c = 0.18a$. Taking advantage of the fact that $Z_e$ is purely imaginary inside the band gaps and purely real inside the passband, in Fig. 5(b), we plot $\text{Im}(Z_e)$ as a function of $\omega$ and $k_p$. Here red, blue and green represent $\text{Im}(Z_e) > 0$, $\text{Im}(Z_e) < 0$ and $\text{Im}(Z_e) = 0$, respectively. Red and blue represent band gaps and green represents passband. The purple triangular region at the top right of Fig. 5(b) is the region where not all the higher order diffraction terms ($g_m \neq 0$) are decaying waves, and in this region, the assumptions to build the layer EMT fail. Close to the purple triangle, the role of higher order diffraction modes become important and the effective parameters from our EMT is not good. Apart from those regions and for the majority of the region under consideration, our EMT gives the salient features of the projected band structure. More importantly, in addition to the projected band structure, our EMT provides additional information for the band gaps. Now, we have two different band gaps, red and blue, corresponding to gaps with $\text{Im}(Z_e) > 0$ and $\text{Im}(Z_e) < 0$, respectively. This classification of band gaps is related to the Zak phases of bulk bands[26, 27] and it has important implication in prediction the existence of surface states and which we will discuss later.

### D. Transmission spectrums compared with full wave simulation

To further check the validity of the layer EMT, we calculate the transmission spectrum (including both amplitude and phase) of a PC slab of finite thickness consisting of eight columns of cylinders inside the vacuum. Here we consider three different incident angles ($\theta$): $\theta = 0°$ for Figs. 6(a) and 6(b), $\theta = 30°$ for Figs. 6(c) and 6(d) and $\theta = 60°$ for Figs. 6(e) and 6(f). Panels (a), (c) and (e) show the amplitude and panels (b), (d) and (f) show the phase. The parameters of the square lattice PC are given by $\varepsilon_c = 10$, $r_c = 0.18a$. Dashed red line and solid blue line show respectively the results calculated with our layer EMT and full wave simulation. Purple represents the region where higher order diffraction waves are not all decaying waves. Our EMT shows perfect agreement with the full wave transmission spectra when the working frequency is not too high or the incident angle is small. We note in particular that both the amplitude and phase are predicted correctly by the layer EMT up to the first gap for a rather high incident angle of 60 degrees. When neither of these two conditions is well satisfied (near the purple region), the roles of the higher order diffraction terms become important, and thus our EMT is not accurate.

## IV Application

We have compared the layer EMT with well-established EMTs like the CPA method, the boundary EMT and full wave simulations. In this section, we will discuss some of the possible applications of our layer EMT. As $Z_e$ describes the surface properties of a PC, it has important implications in predicting the existence of surface states between PCs or between a PC and a homogenous material like the vacuum.

It is well-known that it is difficult to create a surface state localized at the surface of a truncated dielectric PC without "decorating" the top layer of the PC[3]. Here we will use our EMT to explain the reason why the surface state does not form easily at the PC/air interface. While our explanation will be focused on the TM polarization (the electric field parallel to the axis of the cylinders), the argument can be easily extended to the TE polarization. As surface state decays exponentially inside the vacuum, it must be outside the light cone, requiring $k_p > k_0$ where $k_p$ is the wave vector parallel to the surface and $k_0$ is the wave vector inside the vacuum. According to our definition, the $k_p$-dependent effective surface impedance of the vacuum is given by $-E_z/H_y = \sqrt{\mu_0}/\sqrt{(1-k_p^2/k_0^2)\varepsilon_0}$ (the surface is assume to be along the y direction as Fig. 1(a)), which is a negative imaginary number when $k_p > k_0$. The necessary and sufficient condition for the existence of a surface state is

$$\mathrm{Im}\left[Z^R(\omega,k_p)\right] + \mathrm{Im}\left[Z^L(\omega,k_p)\right] = 0, \quad (20)$$

where $Z^R/Z^L$ is the surface impedance of the system on the right/left hand side of the interface. Thus, to create a surface state, the PC must provide a region of $(\omega, k_p)$ where $\mathrm{Im}\left[Z_e(\omega,k_p)\right] > 0$ since the effective surface impedance of a vacuum is a negative imaginary number. According to our EMT, a conventional full photonic band gap (see, for example, band gap II near $\omega a/(2\pi c) = 0.4$ in Fig. 5(b)) does not satisfy this condition. There is indeed a red region where $\mathrm{Im}\left[Z_e(\omega,k_p)\right] > 0$ as shown in Fig. 5(b), gap III near $\omega a/(2\pi c) = 0.45$. However, usually this red region is above the light cone for materials commonly used to make PCs. To illustrate these statements, we increase the permittivity of the cylinder and thus the projected band shift downward globally. After a critical value of the permittivity, the projected gap with $\mathrm{Im}\left[Z_e(\omega,k_p)\right] > 0$ (red region in Fig. 5(b)) becomes partially outside light cone. In Fig. 7, we show one example with surface states existing between a perfect square lattice PC and a vacuum, where the parameters of the cylinders are given by $\varepsilon_c = 45$, $r_c = 0.18a$. The green color marks the passband of the projected band structure of the PC along the interface ($\bar{\Gamma}\bar{X}$) direction and yellow shows the light cone. The surface impedance of the PC inside the band gap near $\omega a/(2\pi c) = 0.31$ satisfies $\mathrm{Im}\left[Z_e(\omega,k_p)\right] > 0$. As said before, there should exist an interface state inside this common gap region which is confirmed (the red line in Fig. 7) with the full wave simulation (We use FDTD [28]). For a comparison, we also solve Eq. (20) numerically using our EMT, and the results are shown with the open blue circles. The blue circles confirm the existence of surface state and can also qualitatively predict the dispersion of the surface state. One thing might be strange at the first sight that the dispersion of surface state predicted with our EMT (open blue circles) terminates before it reaches the bulk projected band. This is because the effects of the higher interlayer diffraction orders become important when $k_p$ becomes larger. This can also be seen by comparing Figs. 5(a) and 5(b), where gap III extends to the BZ boundary in Fig. 5(a) while it closes before reaching the BZ boundary in Fig. 5(b). To recap, our EMT provides an explanation why usually there is no surface state between a dielectric PC and a vacuum.

Besides giving the projected band structures and classifying the band gaps, $\text{Im}\left[Z_e\left(\omega, k_p\right)\right]$ can also be used to predict the dispersions of the surface waves when the approximations we took are well satisfied. In Fig. 8, we study the surface state between a PC and a homogenous media. As the frequency of the surface state depends on the way we truncate the PC, here we terminate the PC so that the outmost unit cell of the PC is a complete unit cell near the boundary as shown in Fig. 8(a) (the distance between the boundary of the homogenous media and the closest cylinder is half the lattice constant along the *x* direction). If the termination of the system is not at exactly the middle point of two columns of cylinders while still not near the cylinders, we only need to add an additional phase due to the shift of boundary relative to our current termination when calculating the dispersion of the surface waves. However, when the termination of the boundary is near the cylinder, then higher order interlayer diffraction terms will become important which also shift the dispersion of the surface wave. The system is periodic along the y direction, and the PC and the homogenous media are both semi-infinite along the x direction. The parameters of the square lattice PC are $\varepsilon_c = 12.5$, $r_c = 0.22a$. In Figs. 8(b)-8(d), we show the interface states inside band gaps I to III of Fig. (5), respectively. Here green represents the projected band of the PC along the $\overline{\Gamma X}$ direction calculated with full wave simulations. The relative permittivity and permeability of the homogenous media are set to be $\varepsilon_2 = 1 - (0.86\pi c/\omega a)^2$, $\mu_2 = 1$ in Fig. 8(b), $\varepsilon_2 = 1, \mu_2 = 1 - (\pi c/\omega a)^2$ in Fig. 8(c) and $\varepsilon_2 = 1$, $\mu_2 = 1 - (0.4\pi c/\omega a)^2$ in Fig. 8(d), respectively. The light yellow regions in Figs. 8(b) and 8(d) represent the passbands of the corresponding homogenous material (light yellow denotes the region where $\text{Im}\left[Z_2(\omega, k_p)\right] = 0$ for this homogenous media). We can solve Eq. (20) numerically and the results are shown in Figs. 8(b)-8(d) with the open blue circles. For comparison, we also give the exact dispersions of the surface waves (the solid red lines) computed with FDTD method[28]. Near the $\Gamma$ point, the EMT predicts the frequencies of surface states quite well; when $k_p$ becomes larger, as the interlayer high-order diffraction terms become important, the blue circles deviate progressively from the red line.

## V Discussion and Extension

Here, we discuss some of the extensions of this kind of EMT and also some possible limitations of this effective medium theory method.

Our layer EMT can apply from 1D to 3D PCs. In this paper, we focus on 2D PCs. Our layer EMT is the same as the retrieval method when considering 1D PCs (See *e.g.* Koschny et al[10]) with a single mirror symmetric unit cell. In 3D, for a nonmagnetic particle, the lowest excitation is dipole mode, and hence the transmission and reflection spectrum can be obtained with a layer of dipole array and the effective parameters are then retrieved also using Eqs. (17)-(19).

In all the examples above we assume a square lattice, here we discuss the case when the cylinder array consists of rectangular lattice ($a \neq b$). In the cases of rectangular lattice, the PCs are intrinsically anisotropic in the *xOy* plane and the effective permittivity and permeability (if they can be defined) become matrixes. As the *x* and *y* directions are the principal axis of the PC under consideration, there are no off-diagonal terms for the permittivity and permeability matrixes. For the TM polarization (the electric field parallel to the axis of the cylinder) under consideration and in the region where the PCs can be described with effective permittivity and permeability, the effective properties of the PCs are described by $\{\mu_x, \mu_y, \varepsilon_z\}$. The surface impedance in Eq. (17) and $n_e$ in Eq. (18) are equivalent to $Z_e = \sqrt{\mu_y / \varepsilon_z}$ and $n_e = \sqrt{\varepsilon_z \mu_y}$, respectively. Note here for rectangular lattice we still need to ignore all the higher interlayer diffraction orders, and this approximation should be better if *a>b*, but will fail faster if *a<b*.

While we have considered the TM polarization, the method can be easily extended to the TE polarization (the magnetic field parallel the axis of the cylinder). In this case, the lowest excitations of cylinder are one magnetic monopole and two in plane (XOY plane) electric dipoles. Because the Maxwell equations without source have the same form if we change $\{\varepsilon, \mu, \vec{E}, \vec{H}\}$ to $\{\mu, \varepsilon, \vec{H}, -\vec{E}\}$, the effective parameters can be obtained from this transformation with no additional effort. The only difference are the excitation strengths of the monopole and dipoles due to the exchange of $\varepsilon$ and

$\mu$.

If we consider the inverted structure, *i.e.*, air cylinders embedded in a dielectric background, the problem becomes different. At low frequency, the energy is concentrated inside the dielectric background instead of the air cylinder. The field distribution in this case cannot be approximately described with only the monopole and dipoles at the cylinder and higher multipoles must be involved. One need to add higher multipoles but then the reflection and transmission spectrum cannot be obtained analytically easily. The proof in Appendix B in this case is still valid as long as we only keep the lowest order diffraction term. There are however, some disadvantages introduced by the dielectric background: the frequency at which higher order diffraction terms are not all decaying wave (purple region in Fig. 5) becomes lower and the effective phase space of our EMT is squeezed. For a similar argument and to reduce the effect of higher order interlayer diffraction terms, we set the boundary of the unit cell inside the lower dielectric material where the higher diffraction orders decay faster. If the dielectric cylinders in the 2D PC are replaced by objects with other shapes, our layer EMT still works as long as the scattering properties of the objects are well described with the monopole and dipole excitations.

## VI. Conclusion

We developed an angular dependent EMT analytically based on the layer KKR method combined with the retrieval method. We assume that the response of the cylinder can be described with the monopole and the dipole excitations and we ignore all the higher order interlayer scattering terms.

This layer EMT works well in the low frequency limit. The method is useful for determining the existence of surface states for photonic crystals without the need to go through large scale calculations. The imaginary part of the effective impedance can be used to predict the projected band structure and it is valid for the majority region of the first BZ as long as our assumptions are still valid. The imaginary part of the surface impedance gives us a classification of band gaps, which is gap with either positive or negative imaginary part of the surface impedance. This gap classification also explains why it is difficult to find surface states between a PC and a vacuum. In the last part of this paper, we show that this layer EMT can be used to find the dispersions of surface states between PCs and homogenous materials.

# ACKNOWLEDGMENTS

This work is supported by Hong Kong RGC through AOE/P-02/12, We would like to thank Professor Z. Q. Zhang for helpful discussions and Mr. Y. Ding for help with the lattice sum.

# APPENDIX

A. EFFECTIVE PARAMETERS FROM RETRIEVAL

Now we will show that $Z_e$ is purely real or purely imaginary and $\cos(n_e k_0 a)$ is purely real if the system has no absorption or gain, which is an important property for an EMT.

From Eqs. (15) and (16), we have

$$t - r = \left(1 - i\tilde{M}_y / b\right) e^{ik_x a}. \quad (A1)$$

From Eq. (11), we have

$$1 - i\tilde{M}_y / b = \frac{ik_x/b + 1/\alpha_M + F_4}{(1/\alpha_M + F_4)}. \quad (A2)$$

Among the regions we consider, all the diffraction terms with nonzero $g_m$ are evanescent waves, so $q_m$ is pure imaginary for $m \geq 1$, and with Eq. (10c),

$$\text{Im}[F_4] = \left(\frac{k_0^2}{8} - \frac{k_x}{2b}\right). \quad (A3)$$

As the cylinder has no gain or loss, $\alpha_M$ contains only the scattering loss, and

$$\text{Im}[1/\alpha_M] = -k_0^2/8. \quad (A4)$$

Combining Eqs. (A3) and (A4), we have

$$\text{Im}[1/\alpha_M + F_4 + ik_x/b] = -\text{Im}[1/\alpha_M + F_4] \quad (A5)$$

So the dominator and numerator on the left-hand side of Eq. (A2) have the same real part and opposite imaginary part, which means that

$$|1 - i\tilde{M}_y/b| = 1, \quad (A6)$$

$$(t-r)(t^* - r^*) = 1. \quad (A7)$$

As the system has no gain or absorption, so from the conservation of the total energy, we have

$$|t|^2 + |r|^2 = 1. \quad (A8)$$

Together with Eq. (A7), we have

$$t^* r + t r^* = 0. \quad (A9)$$

Eq. (A9) is an important intermediate result based on which we can prove our argument that the effective parameters from retrieval are purely real or purely imaginary. One important factor should be noted here that Eq. (A9) is based on the condition that the cylinders are at the center of the unit cell. And when the cylinders are not placed at the center of the unit cell, the effective parameters would not be pure real or pure imaginary even when there is no absorption or gain in the system. In this case, the parameters retrieval from one column of cylinders and those from two columns of cylinders would be different. Then to obtain a set of meaningful effective parameters, one needs several columns of cylinders for retrieval and also proves that the effective parameters converge as the number of columns increasing. Basing on Eq. (A9), it is easy to show that

$Z_e(\omega, k_p)$ is pure real or pure imaginary $\Leftrightarrow$

$$\text{Im}\left[\frac{(r+1)^2 - t^2}{(r-1)^2 - t^2}\right] = 0 \xleftrightarrow{\text{keep denominator real}}$$

$$\text{Im}\left[\left[(r+1)^2 - t^2\right]\left[(r^*-1)^2 - (t^*)^2\right]\right] = 0 \xleftrightarrow{\text{keep only imaginary term}}$$

$$\text{Im}\left[(r-r^*)(1-|r|^2) + (t^2 r^* - r(t^*)^2)\right] = 0 \xleftrightarrow{|t|^2 + |r|^2 = 1}$$

$$\text{Im}\left[(r-r^*)|t|^2 + (t^2 r^* - r(t^*)^2)\right] = 0 \longleftrightarrow$$

$$\text{Im}\left[(t-t^*)(t^* r + t r^*)\right] = 0$$

$\cos(n_e k_0 a)$ is purely real $\Leftrightarrow$

$$(1 + t^2 - r^2)t^* \in \text{Reals} \xleftrightarrow{|t|^2 + |r|^2 = 1}$$

$$(t^* + t) - r(t^* r + t r^*) \in \text{Reals} \Longleftrightarrow$$

$$\text{Im}(t^* r + t r^*) = 0$$

So $Z_e$ is purely real or purely imaginary and $\cos(n_e k_0 a)$ is purely real.

B. EQUIVALENT OF RETRIEVAL FROM ONE COLUMN AND SEVERAL COLUMNS OF CYLINDERS

In this section, we will show that the effective parameters retrieved from one column of cylinder and several columns of cylinders are the same once the unit cell for retrieval has inversion symmetry relative to the center of the unit cell. In Fig. 9, we show a sketch of the idea of the retrieval method: Fig. 9(a) is the real system and Fig. 9(b) is the corresponding equivalent system. The system is periodic along the $y$ direction and here for simplicity we only show the sketch of retrieval from one column of cylinders. Because we ignore all the diffraction terms with $g_m \neq 0$, the relation between $E_L^{+(-)}$ and $E_R^{+(-)}$ in BFig. 1(a) can be described by a transfer matrix, *i.e.*,

$$\begin{pmatrix} E_L^+(x=-a/2) \\ E_L^-(x=-a/2) \end{pmatrix} = M^{1PC}(a) \begin{pmatrix} E_R^+(x=a/2) \\ E_R^-(x=a/2) \end{pmatrix}, \text{(B1)}$$

where $M^{1PC}(a)$ is a $2\times 2$ matrix, the superscript $1PC$ represents one column of cylinder and $a$ is the width of the unit cell. When the unit cell of the real system is replaced by the slab of effective media with the same width, as shown in BFig. 1(b), the relation can be written as

$$\begin{pmatrix} E_L^+(x=-a/2) \\ E_L^-(x=-a/2) \end{pmatrix} = M^{1e}(a) \begin{pmatrix} E_R^+(x=a/2) \\ E_R^-(x=a/2) \end{pmatrix}, \text{(B2)}$$

where $M^{1e}(a)$ is still a $2\times 2$ matrix, the superscript $1e$ represents effective media from one column of cylinders, and $a$ is the width of the slab. From the point of retrieval method, we only need the reflection and transmission coefficients to be the same, namely,

$$M_{11}^{1PC} = M_{11}^{1e}, \quad M_{21}^{1PC} = M_{21}^{1e}. \text{(B3)}$$

For a unit cell with inversion symmetry relative to the cell of unit cell, one can have

$$M_{21}^{1PC} = -M_{12}^{1PC}. \text{(B4)}$$

which means that,

$$M_{12}^{1PC} = M_{12}^{1e}. \text{(B5)}$$

And then the transfer matrix in Eq. (B1) and Eq. (B2) are exactly the same, i.e.,

$$M^{1PC}(a) = M^{1e}(a). \text{(B6)}$$

So if we want to use the retrieval method with two columns of cylinders, the transfer matrix reads

$$M^{2PC}(2a) = M^{1PC}(a)M^{1PC}(a) = M^{1e}(a)M^{1e}(a) = M^{1e}(2a). \text{(B7)}$$

As the inversion symmetry still exists, from the retrieval method, we still have

$$M^{2PC}(2a) = M^{2e}(2a). \text{(B8)}$$

Then, combined with Eq. (B7), we have

$$M^{2e}(2a) = M^{1e}(2a). \text{(B9)}$$

So the effective parameters for 1 and 2 unit cells are the same. For retrieval with more unit cells, the same argument can be followed as above, so with the inversion symmetry, the effective parameters retrieved from one column of cylinders and several columns of cylinders are the same. For simplicity, we only need to consider one column of cylinders for retrieval.

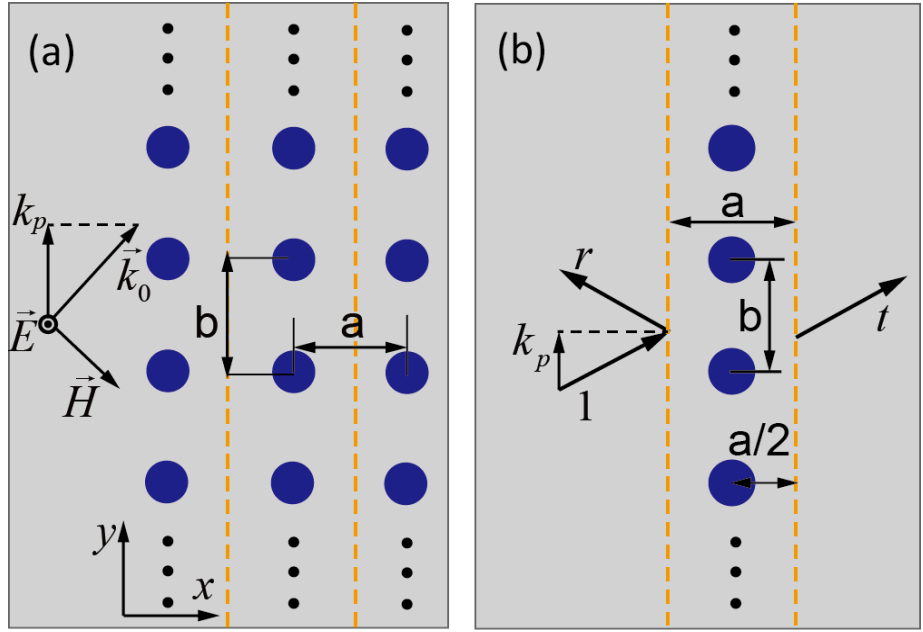

FIG.1. (Color online) (a) System consisting of identical cylinders (blue disks) embeded in a vacuum, the lattice constant is a (b) along the *x*-(*y*-) direction. The E field of the incident plane wave is along the axis of the cylinders (the *z*-direction) and the component of the wave vector along the *y*-direction is given by $k_p$. This system can be treated as a stack of layers with boundaries marked by the dashed orange lines. (b) A plane wave with unit amplitude impinges on a single column of cylinders, and the transmission and reflection coefficients are given by $t$ and $r$, respectively. The virtual boundaries of the column (marked with dashed orange lines) are chosen such that the cylinders are at the center of the column.

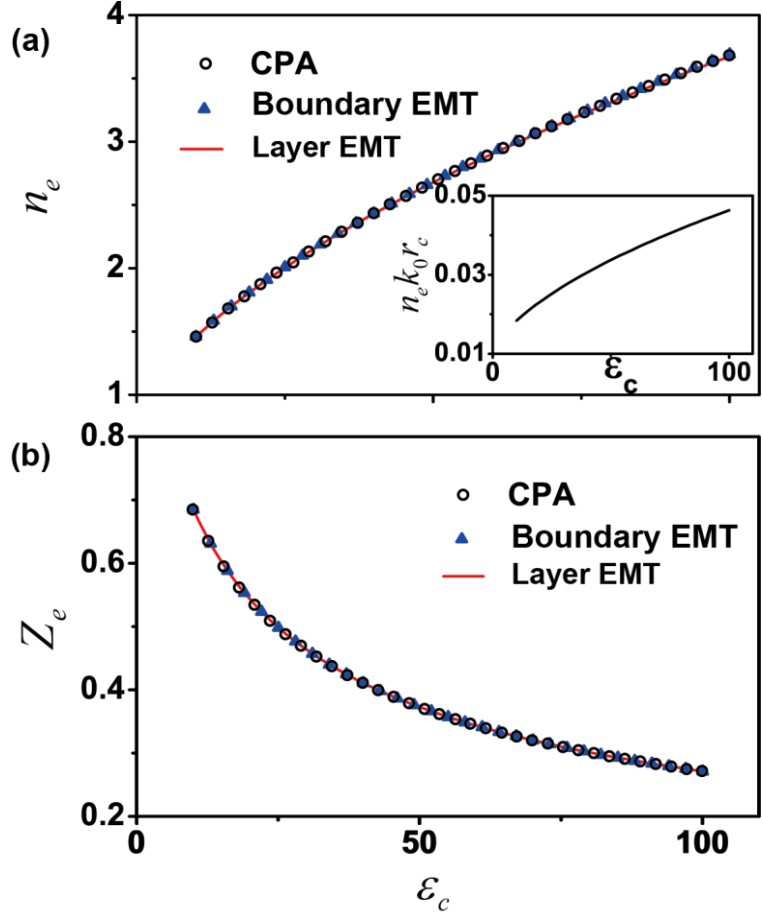

FIG. 2. (Color online) Effective refractive index (a) and effective relative impedance (b) as a function of the relative permittivity of the cylinders ($\varepsilon_c$). The angular frequency and the radius of the cylinder ($r_c$) are $0.02\pi c/a$ and $0.2a$ respectively, where $c$ is the speed of light in a vacuum and $a$ is the length of the square unit cell. The open black circles, solid blue triangles and solid red line show the results calculated using the conventional CPA-based method, the boundary effective media theory (BEMT) and our layer EMT, respectively. The inset in (a) shows $n_e k_0 r_c$ as a function of $\varepsilon_c$.

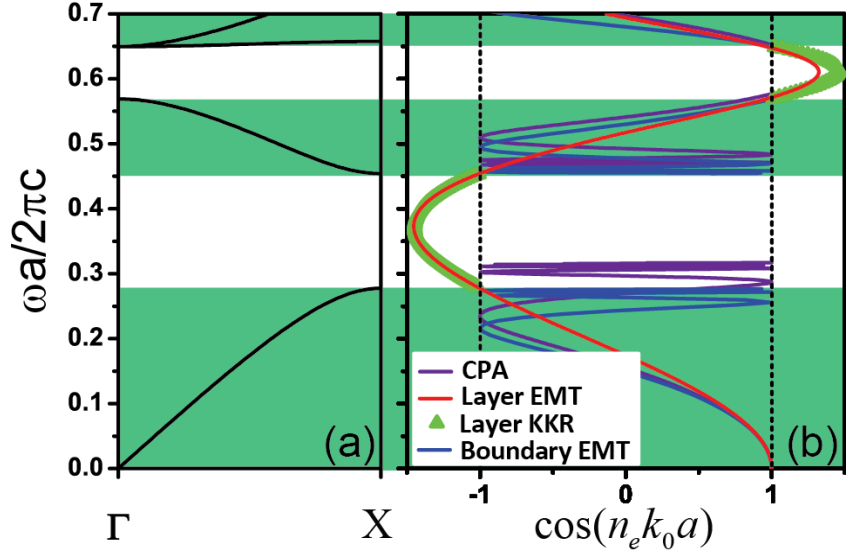

FIG. 3. (Color online) (a) Band dispersion along the $\Gamma X$ direction. (b) Purple line, red line, green triangles and blue line show $\cos(n_e k_0 a)$ calculated with the CPA method, our layer effective media theory (EMT), the layer KKR method and the boundary EMT, respectively. The dashed vertical black lines enclose the region where $|\cos(n_e k_0 a)| \leq 1$ which corresponds to the passband region. In both (a) and (b), green and white backgrounds represent the passband and gap region, respectively. The relative permittivity and the radius of the dielectric cylinder are $\varepsilon_c = 10$ and $r_c = 0.18a$, respectively, where $a$ is the length of the square unit cell.

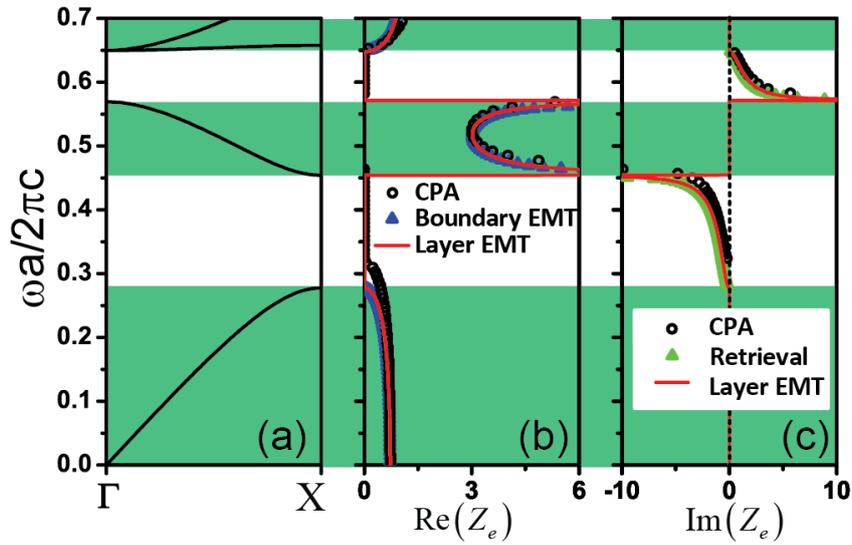

FIG. 4. (Color online) (a) Band dispersion along the $\Gamma X$ direction. (b) Open black circles, solid blue triangles and red line show the real part of the effective impedance ($\text{Re}(Z_e)$) calculated using the CPA method, the boundary effective media theory (BEMT) and our layer EMT, respectively. (c) Open black circles, solid blue triangles and red line show the imaginary part of the effective impedance ($\text{Im}(Z_e)$) calculated using the CPA method, the retrieval method and our layer EMT, respectively. The dashed vertical black line in (c) represents $\text{Im}(Z_e)=0$. In (a)-(c), green and white backgrounds represent the passband and gap regions, respectively. The relative permittivity and the radius of the dielectric cylinder are $\varepsilon_c =10$ and $r_c =0.18a$, respectively, where $a$ is the length of the square unit cell.

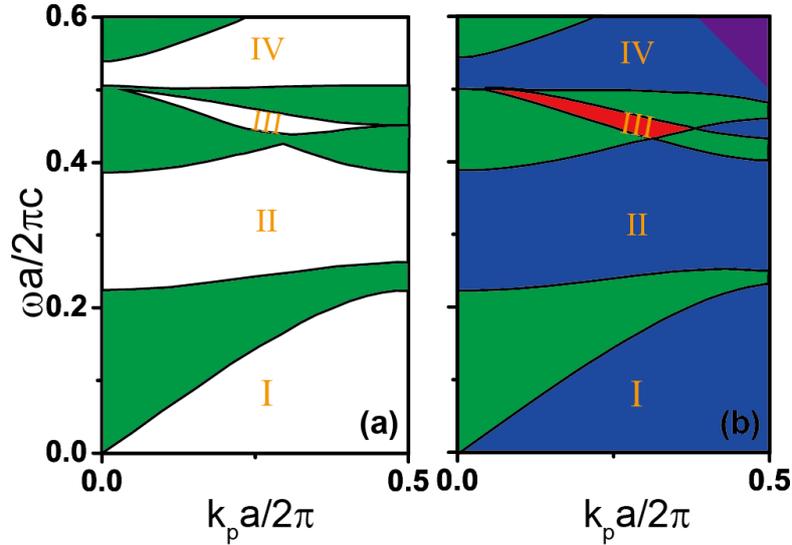

FIG. 5. (Color online) (a) Projected band along the interface direction ($\overline{\Gamma}\overline{X}$) of the PC with parameters $\varepsilon_c =12.5$ and $r_c =0.22a$ for square array of cylinders embedded in a vacuum, with green representing pass bands, and white representing band gaps. (b) Imaginary part of the effective surface impedance for the same PC. Blue for $\text{Im}[Z_e(\omega,k_p)]<0$, green for $\text{Im}[Z_e(\omega,k_p)]=0$ and red for $\text{Im}[Z_e(\omega,k_p)]>0$. The purple triangular region in (b) represents the region where higher order diffraction terms are not all decaying waves. The numbers of bandgaps are labeled in orange.

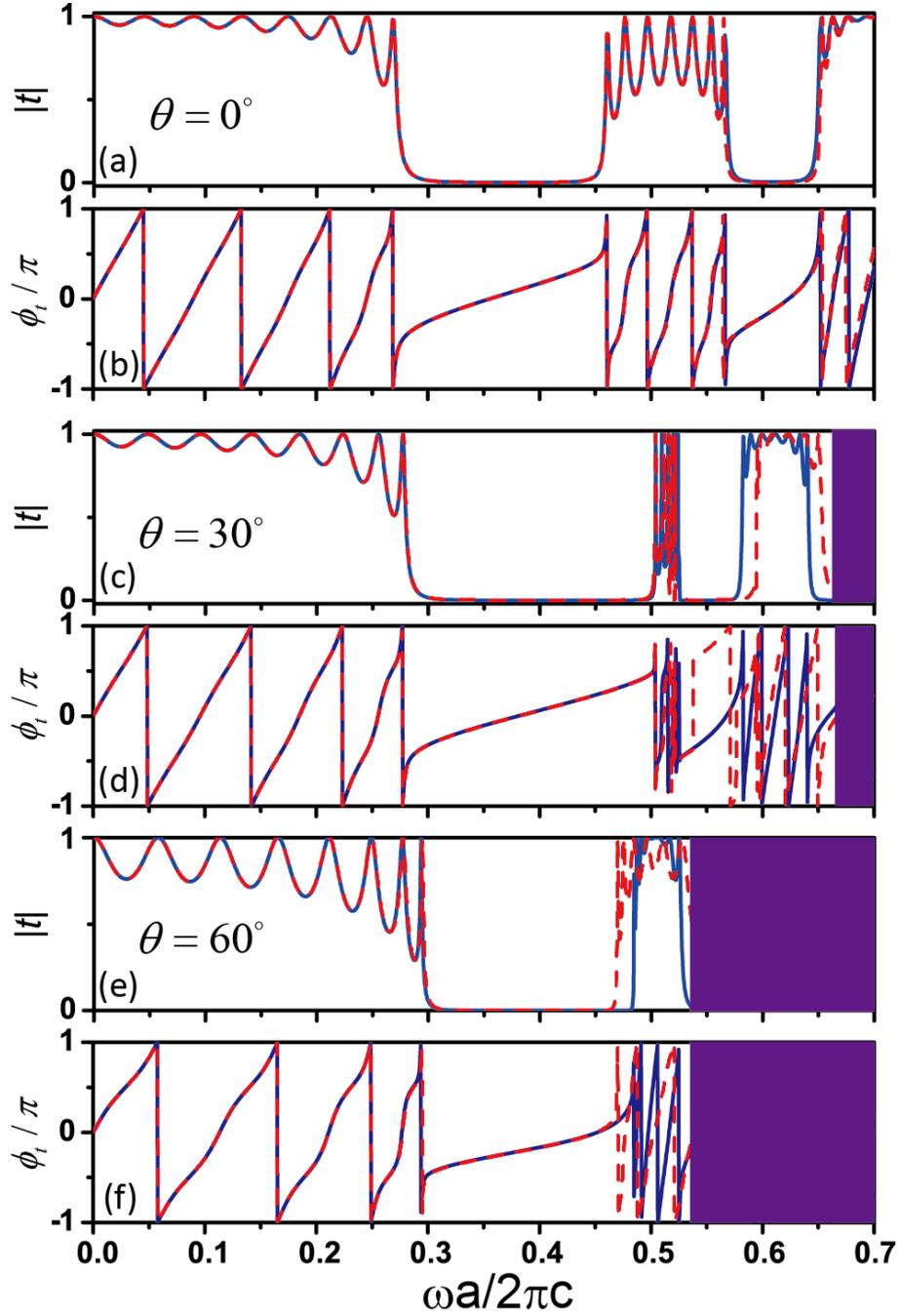

FIG. 6. (Color online) Amplitude and phase of the transmitted waves through eight columns of cylinders with the incident angles of incident waves given by $\theta = 0°$ for (a) and (b), $\theta = 30°$ for (c) and (d), and $\theta = 60°$ for (e) and (f). The parameters of the PC used are given by $\varepsilon_c = 10$, $r_c = 0.18a$ and square lattice. The dashed red line and solid blue line show the results calculated with our effective media theory and full wave

simulation, respectively. Purple rectangular covers the region where higher order diffraction terms are not all decaying waves.

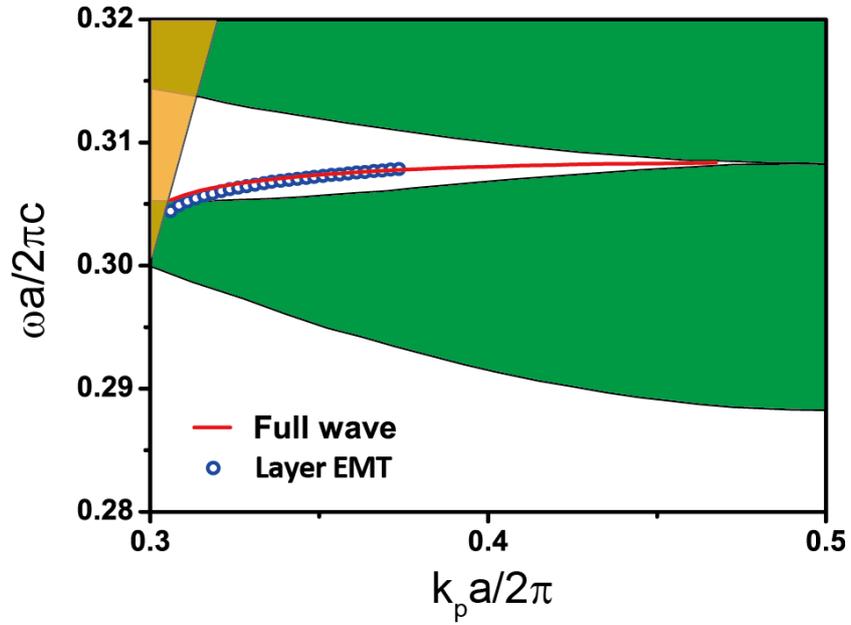

FIG. 7. (Color online) The surface wave between a PC and a vacuum. The relative permittivity and the radius of the cylinder in the PC are $\varepsilon_c = 45$ and $r_c = 0.18a$ respectively, where $a$ is the length of the square unit cell. Green represents the projected band structure of the PC along the interface ($\overline{\Gamma}\overline{X}$) direction calculated using the full wave simulation, yellow represents the light cone of a vacuum and white represents the common band gap. The red line calculated using full wave simulation which shows the dispersion of the surface states inside this system and the blue circles are obtained with layer effective media theory.

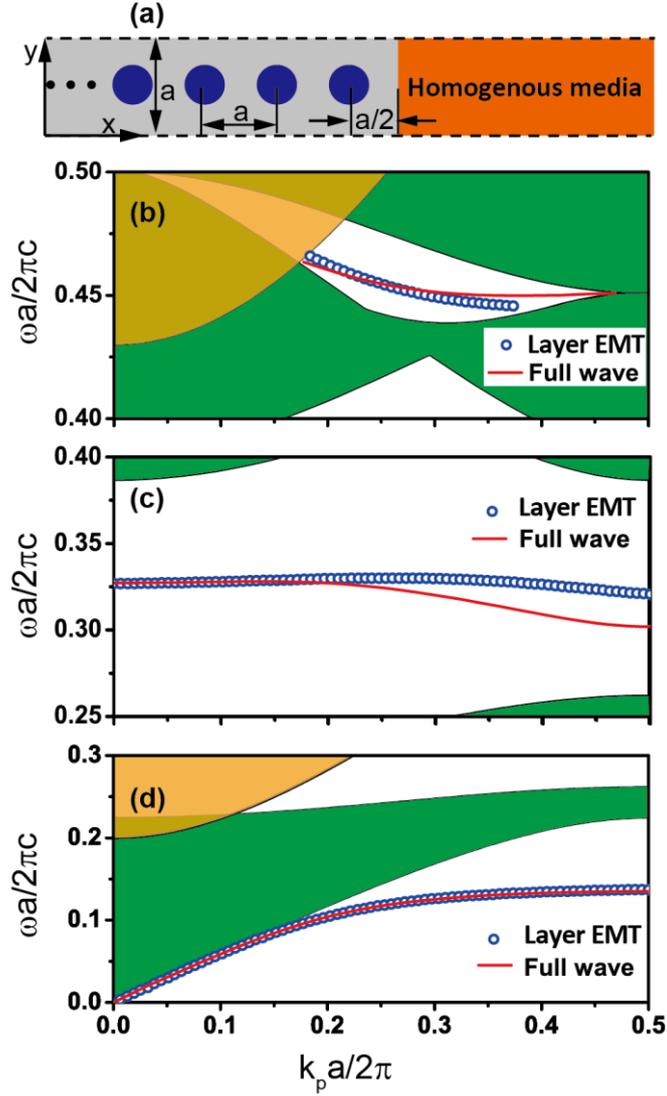

FIG. 8. (Color online) (a) Construction of an interface between a PC and a homogenous media. The system is periodic along the *y* direction, and the PC and the homogenous media are both semi-infinite along the *x* direction. (b)-(d) The interface states and the projected band structures between a PC and different homogenous medias. The parameters of the PC are $\varepsilon_c = 12.5$ and $r_c = 0.22a$, where $a$ is the length of square unit cell. The relative permittivity and permeability of the homogenous medias are given by $\varepsilon_2 = 1 - (0.86\pi c/\omega a)^2$, $\mu_2 = 1$ for (b), $\varepsilon_2 = 1$, $\mu_2 = 1 - (\pi c/\omega a)^2$ for (c) and $\varepsilon_2 = 1$, $\mu_2 = 1 - (0.4\pi c/\omega a)^2$ in (d), respectively. Green represents the passband of the PC along the interface ($\overline{\Gamma X}$) direction, yellow represents the region where the wave is propagating wave inside the homogenous media, and the white represents common band gap region. Red lines show the dispersions of the surface states calculated

using full wave simulation and the open blue circles are calculated using our effective media theory.

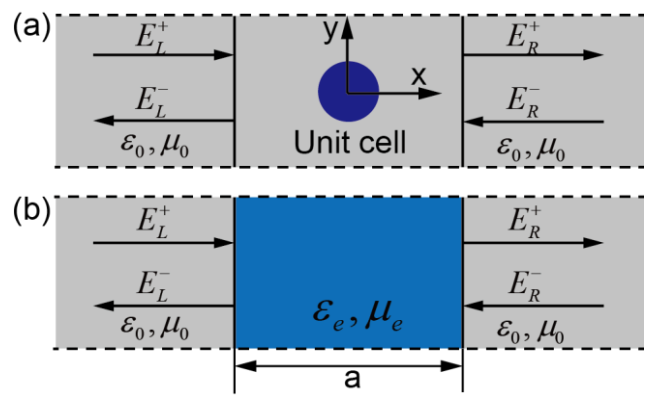

FIG. 9. (Color online) Sketch of the idea of the retrieval method, (a) the real system and (b) the effective system. Both systems are periodic along the *y* direction.